\documentclass[12pt,preprint]{aastex}
\usepackage{amsmath, amsthm, amssymb}
\newcommand{\rsun}{R_{\odot}}

\begin{document}

\title{The effect of activity-related meridional flow modulation on the
strength of the solar polar magnetic field}
\author{J. Jiang\altaffilmark{1}, E. I\c{s}\i k\altaffilmark{2},
        R.H. Cameron\altaffilmark{1}, D. Schmitt\altaffilmark{1}
        \& M. Sch{\"u}ssler\altaffilmark{1}}

\altaffiltext{1}{Max-Planck-Institut f\"ur Sonnensystemforschung,
   37191 Katlenburg-Lindau, Germany}

\altaffiltext{2}{Department of Physics, Faculty of Science \& Letters,
   \.Istanbul K\"ult\"ur University, Atak\"oy Campus,
   Bak\i rk\"oy 34156, \.Istanbul, Turkey}

\email{jiang@mps.mpg.de}

\begin{abstract}
We studied the effect of the perturbation of the meridional flow in the
activity belts detected by local helioseismology on the development and
strength of the surface magnetic field at the polar caps. We carried out
simulations of synthetic solar cycles with a flux transport model, which
follows the cyclic evolution of the surface field determined by flux
emergence and advective transport by near-surface flows. In each
hemisphere, an axisymmetric band of latitudinal flows converging towards
the central latitude of the activity belt was superposed onto the
background poleward meridional flow. The overall effect of the flow
perturbation is to reduce the latitude separation of the magnetic
polarities of a bipolar magnetic region and thus diminish its
contribution to the polar field. As a result, the polar field maximum
reached around cycle activity minimum is weakened by the presence of the
meridional flow perturbation. For a flow perturbation consistent with
helioseismic observations, the polar field is reduced by about 18\%
compared to the case without inflows. If the amplitude of the
flow perturbation depends on the cycle strength, its effect on the polar
field provides a nonlinearity that could contribute to limiting the
amplitude of a Babcock-Leighton type dynamo.
\end{abstract}

\keywords{Sun: activity, Sun: magnetic fields, Sun: meridional circulation}

\section{Introduction}
Surface flux transport models treat the evolution of the large-scale
magnetic field on the surface of the Sun
\citep[e.g.,][]{Wang:etal:1989b, Schrijver:2001, Mackay02,
Baumann04}. In such models, the evolution of the radial magnetic field
at the solar surface is governed by the emergence of new flux in the
form of bipolar magnetic regions and by advective transport through
large-scale flows (differential rotation, meridional circulation) and
supergranular turbulent diffusion. The well-known cyclic variations of
differential rotation in the form of zonal flows
\citep[e.g.,][]{Howard:Labonte:1980, Howe:etal:2006b} so far have not
been considered in flux transport simulations.  On the other hand,
variations in the large-scale meridional flow \citep{Komm93,
Basu03,Hathaway:Rightmire:2010} have been considered in flux-transport
simulations by assuming cycle-to-cycle changes in the overall amplitude
of the flow \citep{Wang02_b, Dikpati04, Wang:etal:2009}.

Another cycle-related modulation of the surface flow field is the
modulation of the axisymmetric component of the meridional flow in the
form of bands of latitudinal velocity centered on the dominant latitudes
of magnetic activity, first detected at depths greater than 20~Mm \citep
{Chou:Dai:2001, Beck:etal:2002}. In the case of near-surface flows, the
residual meridional flow velocities (after subtraction of the mean flow)
during cycle 23 were of the order of $\pm $3$-$5~m$\,$s$^{-1}$ and
converge toward the dominant latitudes of magnetic activity while
migrating towards the equator in parallel to the activity belts
\citep{Gizon08, Gonzalez08, Gonzalez:etal:2010}.  These flows are
probably related to the meridional motions of sunspots and pores
\citep[][and references therein]{Ribes:Bonnefond:1990} and other
magnetic features \citep{Komm:1994, Meunier:1999}. The cumulative effect
of the near-surface horizontal flows converging towards active regions
\citep[e.g.,][]{Haber:etal:2004, Hindman:etal:2004} appear to contribute
to the axisymmetric meridional flow perturbation \citep{Gizon:2004,
Gonzalez08}, but there is evidence that at least part of this
perturbation is unrelated to surface activity
\citep{Gonzalez:etal:2010}.

The effect of the near-surface inflows on the evolution of single
active regions was recently studied by \citet{De_Rosa06}.
Considering results obtained with a surface flux transport model
\citep[cf.][]{Schrijver:2001}, these authors find that inflows of
the order of $\sim10$~m$\,$s$^{-1}$ significantly affect the
dispersal of magnetic flux from an isolated active region. These
results indicate that the axisymmetric meridional flow perturbations
associated with the activity belts could also affect the evolution
of the solar surface field on a global scale. Particularly
interesting in this connection is the effect on the polar field
strength, which is an important source of the heliospheric field and
also plays a significant role in Babcock-Leighton-type dynamo
models. Here we present results of solar-cycle simulations using the
flux transport code of \citet{Baumann04}, including axisymmetric
bands of converging latitudinal flows centered on the migrating
activity belts.   The aim of this work is to study the general
effect of these flows on the evolution of the solar surface field,
and particularly on the strength of the polar field. This is an
exploratory study focussing on understanding the physical
mechanisms; we do not intend to reproduce any actual solar data. We
need not consider the zonal flows in this study because the buildup
of magnetic field to the Sun's poles is dominated by the latitude
separation of the polarities of a bipolar magnetic region and thus
essentially is an axisymmetric problem \citep{Cameron07}; zonal
flows \citep[and differential rotation in general,
see][]{Leighton:1964} have no effect on the amount of signed flux
reaching the poles.

This paper is organized as follows. The flux transport model is
described in Section~\ref{sec:model}. The relevant effects of the
latitudinal flow bands on the surface flux evolution are illustrated
with simulations of single bipolar regions in
Section~\ref{sec:bipoles}. The results of full solar-cycle simulations
are presented in Section~\ref{sec:cycles}, which includes a study of the
dependence of the polar field on various model parameters. The
implication of our results are discussed in Section~\ref{sec:disc}.

\section{Flux-transport model}
\label{sec:model}

The induction equation considered in our flux transport model is given
by \citep[for details see][]{Baumann04, Jiang09, Jiang10}
\begin{eqnarray}
\label{eqn:SFT}
\nonumber\frac{\partial B}{\partial t}=& & -\Omega(\lambda,t)
                       \frac{\partial B}{\partial \phi}
         - \frac{1}{\rsun \cos\lambda}
              \frac{\partial}{\partial \lambda}[v(\lambda,t)
         B \cos \lambda] \\ \noalign{\vskip 2mm}
& & +\eta_{H} \left[\frac{1}{\rsun^2 \cos{\lambda}}
                \frac{\partial}{\partial \lambda}\left(\cos\lambda
          \frac{\partial B}{\partial \lambda}\right) +
     \frac{1}{\rsun^2 \cos^2{\lambda}}\frac{\partial^2 B}{\partial
     \phi^2}\right]\\ \noalign{\vskip 2mm} \nonumber
& & + S(\lambda,\phi,t) +D(\eta_r),
\end{eqnarray}
where $\phi$ and $\lambda$ are longitude and latitude, respectively,
$B$ is the radial component of the magnetic field, $\Omega$ is the
rotational velocity, $v$ is the meridional flow velocity, $\eta_{H}$
is the turbulent surface diffusivity due to the random granular and
supergranular velocity field, $S$ is a source term which describes
the emergence of new flux, and the term $D(\eta_r)$ models the
radial diffusion of the field \citep{Baumann06} with the diffusivity
parameter set to $\eta_r=100$~km$^2$s$^{-1}$. We use the synodic
rotation rate $\Omega=13.38-2.30 \sin^2 \lambda -1.62 \sin^4
\lambda$ (in degrees per day) { determined by \citet{Snodgrass83}
and take $\eta_H=600$~km$^2$s$^{-1}$.

The meridional flow velocity consists of a background flow plus a
perturbation, $\Delta v(\lambda,t)$, representing axisymmetric bands of
converging latitudinal flow (one per hemisphere), viz.
\begin{equation}
v(\lambda,t)=\begin{cases} v_{\rm m} \sin(2.4\lambda)+
   \Delta v(\lambda,t) & \text{for\ } \vert\lambda\vert \le 75^{\circ}\\
   0 & \text{otherwise,}
\end{cases}
\label{eqn:mer}
\end{equation}
where $v_{\rm m}=11$~m$\,$s$^{-1}$ and
\begin{eqnarray}
\Delta v(\lambda,t)=  \left\{ \begin{array}{l l}
  v_0 \sin\left[(\lambda-\lambda_{\rm c}(t))/\Delta\lambda_\upsilon
  \right]& \text{for\ } {-180}^{\circ}\le (\lambda -\lambda_{\rm c}(t))/
  \Delta\lambda_\upsilon < 180^{\circ} \\
          & \\ 0&  \mathrm{otherwise.}
\end{array} \right.
\label{eqn:bands}
\end{eqnarray}
The bands of perturbed meridional flow are characterized by their
velocity amplitude, $v_0$, their width, $\Delta\lambda_\upsilon$, and
their central latitude, $\lambda_{\rm c}$. The equatorward migration of
the bands in the course of the solar cycle is represented by the time
dependence of $\lambda_{\rm c}$ (see Section~\ref{subsec:cycle_param}).
Note that Equation~(\ref{eqn:bands}) describes one band, its counterpart
on the other hemisphere is obtained by changing $\lambda_{\rm c} \to
-\lambda_{\rm c}$. For sufficiently small central latitudes, the two
bands can overlap and the corresponding velocities are added.

\section{Evolution of single bipolar magnetic regions}
\label{sec:bipoles}

In order to illustrate the effect of the meridional flow perturbation on
the latitudinal flux transport as the source of the polar field, we
first study a single bipolar magnetic region (BMR). The temporal
evolution of the corresponding surface flux depends on the relative
position of the bands of latitudinal flow perturbation and the emergence
latitude. We consider the evolution of a BMR that emerges at $t=0$ at a
latitude of $15^\circ$ on the northern hemisphere under the influence of
four different meridional flow patterns (see Figure~\ref{fig:flow})
described by Eqs.~(\ref{eqn:mer}) and (\ref{eqn:bands}). The initial flux
distribution of the BMR is chosen following the approach of
\citet{Baumann04}.

Snapshots of the surface distribution of the magnetic field are shown in
Figure~\ref{fig:bip_snaps}. The four cases shown correspond to no flow
perturbation (top row) and to converging flow bands centered on
different latitudes $\lambda_{\rm c}$. The corresponding time evolution
of the polar fields is shown in Figure~\ref{fig:bip_polar}.  When the
flow perturbation is centered equatorward of the BMR emergence latitude
($\lambda_{\rm c}=\pm5^\circ$, second row in
Figure~\ref{fig:bip_snaps}), the overlap of the flow perturbations from
both hemispheres (see blue curve in Figure~\ref{fig:flow}) has the
consequence that preceding and following polarities of the BMR
experience an increased latitude separation: the leading polarity is
advected toward the equator while the following polarity is less
affected. As a consequence, the latitudinal separation between preceding
and following polarity increases, so that the polar
field becomes stronger in comparison to the case without flow
perturbation. The opposite effect results in the case $\lambda_{\rm
c}=\pm15^\circ$ (third row in Figure~\ref{fig:bip_snaps}, red curves in
Figs.~\ref{fig:flow} and \ref{fig:bip_polar}): now the latitudinal
gradient of the meridional flow at the emergence location is such that
the two polarities are now advected towards each other, thus reducing
the azimuthally averaged field and, consequently, the flux reaching the
pole. In the third case ($\lambda_{\rm c}=\pm25^\circ$, fourth row in
Figure~\ref{fig:bip_snaps}, green curves in Figs.~\ref{fig:flow} and
\ref{fig:bip_polar}), there are two opposing effects: the meridional
flow gradient near the emergence location tends to separate the
polarities while the following polarity experiences an overall decrease
of its poleward advection, thus tending to reduce the azimuthally
averaged field. The net effect is a slight reduction of the contribution
to the polar field.

These results show that the emergence location of a BMR relative to
the position of the bands of flow perturbation is important for its
effect on the development of the polar field.  Note that the
(axisymmetric) meridional flow perturbation considered here results
from the cumulative effect of the individual inflows. A given active
region (which can appear anywhere in the activity belt) therefore
experiences the superposition of these inflows, which needs not
necessarily be centered on this active region.

\section{Simulation of activity cycles}
\label{sec:cycles}

\subsection{Cycle parameters}
\label{subsec:cycle_param}
As next step, we consider sequences of simulated activity cycles by
periodically varying the number of BMRs that appear on the surface.  The
emerging BMRs have a tilt angle of half their emergence latitude,
follow Hale's polarity rules, and are introduced in activity belts that
migrate toward the equator. The BMR area, $A$, follows the distribution
$n(A)\propto A^{-2}$ derived from observations
\citep{Schrijver:Harvey:1994}.  The number of BMRs emerging during the
$i^{\rm th}$ cycle is taken to vary proportional to a Gaussian time
profile, viz.
\begin{eqnarray}
     n_i(t)&\propto&\left\{ \begin{array}{l l}
     \exp\{-[(t-t_i+6.5)/3.25]^2\} & 0 \le (t-t_i) \le13 \\
        0 & \mathrm{otherwise} \end{array} \right.
\label{eqn:bmrs}
\end{eqnarray}
where $t_i=11\times i$ is the starting time of the $i^{\rm th}$ cycle and
all times are in years. With new cycles starting every 11~years and
having a duration of 13 years we thus take into account the overlap of
solar activity cycles.  The emergence of new BMRs occurs randomly with a
Gaussian distribution of half-width $\Delta\lambda_B$ about the central
latitudes of the activity belts, $\lambda_\pm$, which migrate
equatorward according to
\begin{eqnarray}
\lambda_{\pm}(t)&=&\pm[\lambda_0-(\lambda_0-8^\circ)(t-t_i)/13],
\end{eqnarray}
so that the belts progress from their starting latitudes,
$\pm\lambda_0$, to $\pm8^\circ$ in the course of 13 years.  The resulting
emergence pattern of new BMRs (butterfly diagram) for
$\lambda_0=25^\circ$ and $\Delta\lambda_B=6^\circ$ is shown in
Figure~\ref{fig:bf}.

The latitudinal bands of the meridional flow perturbation move in
parallel to the active region belts, their central latitudes (on both
hemispheres), $\pm\lambda_c$, coinciding with the centers of the
corresponding activity belts, $\lambda_\pm$. We do not assume an overlap
of the meridional flow perturbations from consecutive cycles; therefore,
we include the flow perturbation only for 11 years, starting from the
third year of each 13-year cycle. Since the early flux emergence at mid
latitudes affects the polar field only little, this assumption does not
influence the results in a significant way (see also
Section~\ref{subsec:cyc_dep}).

\subsection{Dependence on the flow perturbation parameters}

Figure~\ref{fig:flow_amp} shows the cyclic variation of the polar fields
for three values of the flow perturbation amplitude: $v_0= (0,\, 5,\,
10)$~m$\,$s$^{-1}$. The latitudinal width of the bands was kept
fixed at $\Delta\lambda_v=15^\circ$ and the activity belt parameters
were $\lambda_0=25^\circ$ and $\Delta\lambda_B=6^\circ$.  As already
suggested by the results of the study of single BMRs shown in
Section~\ref{sec:bipoles}, we find that the net effect of the flow
perturbation on BMRs emerging in an extended activity belt is a
reduction of the polar field amplitudes. The effect becomes more
pronounced with increasing flow perturbation amplitude.

On a more quantitative level, the dependence of the mean polar field
amplitude (averages over three consecutive cycles)%
\footnote{We omit the first two simulated cycles from the analysis as
  these could be affected by the arbitrary initial magnetic field.
  With $\eta_r=100$~km$^2$s$^{-1}$, the $e$-folding time of the magnetic
  field in the absence of sources (and thus the `memory' of the system)
  is of about 5 years.}
on the width, $\Delta\lambda_v$, and the amplitude, $v_0$, of the flow
perturbation is given in Table~1. The numbers in parentheses give the
percentage change of the polar field with respect to the case with
unperturbed meridional flow.  In all cases we find a reduction of the
polar field. For parameters roughly corresponding to the
helioseismic results $(v_0=5~$m$\,$s$^{-1}$,
$\Delta\lambda_v=15^\circ)$, the flow perturbation leads to a reduction
of the polar field amplitude by about 18\% with respect to the same case
but without flow perturbation. Apart from the reduction becoming more
pronounced with increasing perturbation amplitude, it also is stronger
for bigger $\Delta\lambda_v$, i.e., for wider bands of perturbed
flow. This is plausible because wider flow bands affect a larger
proportion of the BMRs emerging in the activity belts and, at the same
time, influence latitudinal flux advection for a longer time. In
  all cases, the evolution of the total unsigned surface flux is almost
  unaffected by the presence of the flow perturbation.

\subsection{Dependence on the activity belt parameters}

Keeping the parameters of the meridional flow perturbation fixed at
values of $v_0=5$~m$\,$s$^{-1}$ and $\Delta\lambda_v=15^\circ$, we also
considered the dependence of the polar field amplitude on the starting
latitude, $\lambda_0$, and the width, $\Delta\lambda_B$, of the activity
belt. The results are summarized in Table~2. As already suggested by the
results of Section~\ref{sec:bipoles}, the biggest effect on the polar
field occurs when BMRs always emerge near the center (latitude of
convergence) of the bands of perturbed flow, i.e., for
$\Delta\lambda_B=0^\circ$.  The flow perturbation then always tends to
decrease the latitude extent of the BMR and thus reduces its azimuthally
averaged field. The broader the activity belt (relative to the band of
perturbed flow), the smaller is the effect on the polar field.  On the
other hand, for a given activity belt width, the variation of the
starting latitude, $\lambda_0$, of the activity belt in a cycle does not
significantly change the effect of the flow perturbation. {\bf In most
cases,} there is a tendency for the polar field amplitude to decrease
with increasing $\lambda_0$. This result may have implications for the
nonlinear limitation of a Babcock-Leighton dynamo as further discussed
in Section~\ref{sec:disc}.

\subsection{Time-dependent flow perturbation amplitude}
\label{subsec:cyc_dep}

The observations based on helioseismology indicate that the
amplitude of the axisymmetric flow perturbation peaks around the
maximum of magnetic activity \citep{Gonzalez:etal:2010}. We
therefore also considered the effect of a temporal variation of the
flow perturbation in parallel to the activity level. To this end, we
modulated the perturbation amplitude, $v_0$, with the same time
profile as that assumed for the number of emerging BMRs given by
Equation~(\ref{eqn:bmrs}), so that the maximum speed is reached at
activity maximum. With the previously used parameters for the flow
perturbation ($\upsilon_0=5\,\rm{m\,s^{-1}}$,
$\Delta\lambda_{\upsilon}=15^\circ$) and for the width of the
activity belt ($\Delta\lambda_B=6^\circ$), this results in a polar
field with an amplitude (three-cycle average) of 5.66~G, about 2.5\%
higher than the value of 5.52~G found for constant flow perturbation
amplitude. The effect of the time variation is somewhat stronger if
we assume zero spread of the activity belt ($\Delta\lambda_B=0$); in
this case we obtain a polar field of 5.22~G, which is about 6\%
higher than the corresponding value of 4.92~G for constant flow
perturbation amplitude. This is to be expected since the polar field
is dominated by the trans-equatorial transport (or cancellation) of
leading-polarity flux; therefore, in the case of very narrow
activity belts, the late phase of a cycle with flux emergence near
the equator contributes more strongly to the strength of the polar
field \citep{Cameron07}.

Altogether, the effect of a temporal variation of the inflow
amplitude is found to be rather small.  Since the temporal
modulation strongly reduces the flow perturbation during the rise
and decay phases of a cycle, this result implies that the influence
of the meridional flow perturbation on the polar field is dominated
by the period around activity maximum.

\section{Discussion and conclusion}
\label{sec:disc}

The results presented here show that the observed cycle-related
meridional flow perturbations in the form of bands migrating with the
activity belts decrease the strength of the polar fields resulting from
the latitudinal transport of surface flux. For a flow perturbation
corresponding to the helioseismic observations, this reduction amounts
to about 18\% with respect to the case without flow perturbation. This
indicates that these effects should be taken into account in surface
flux transport simulations aiming at a quantitative pre- or postdiction
of the polar field strength.

It is doubtful whether this kind of flow perturbation could have
significantly contributed to the low polar polar field strength during
the activity minimum between solar cycles 23 and 24
\citep[e.g.,][]{Schrijver:Liu:2008} as compared to previous minima: the
perturbation is probably present during every cycle, so that only an
{\em increase} of the perturbation in cycle 23 compared to its amplitude
in previous cycles would contribute to a comparatively weaker polar
field during the recent minimum. In any case, other effects must have
been affecting the polar field in addition since the observed reduction
by nearly a factor of 2 exceeds the decrease that could be caused by the
flow perturbation considered here.

The observed variation of the flow perturbation amplitude during the
activity cycle \citep{Gonzalez:etal:2010} and the probable contribution of
the near-surface inflows toward active regions to the driving of the
perturbations \citep {Gizon08} suggest that the amplitude of the flow
perturbation should increase with cycle strength. According to our
results, this would lead to a stronger reduction of the polar field
built up during cycles of higher activity. Since, in the framework of a
Babcock-Leighton dynamo, the polar field is a measure of the poloidal
field providing the basis for the toroidal field in the subsequent
cycle, the meridional flow perturbation is potentially important for the
nonlinear modulation and limitation of the cycle amplitude.
Furthermore, we have also seen that the polar field decreases for
increasing starting latitude, $\lambda_0$, of the activity belt at the
beginning of a cycle. Since stronger cycles typically have higher values
of $\lambda_0$ \citep{Solanki:etal:2008}, this relationship would
strengthen the nonlinear effect of the flow perturbation in the
subsequent cycle amplitude.

We conclude that, in addition to global variations of the meridional
flow speed \citep{Wang02_b, Wang02}, the cyclic perturbation of the
meridional flow by converging bands migrating with activity belts has an
appreciable effect on the build-up of the magnetic field at the polar
caps. Its relation to the strength of a cycle means that the flow
perturbation could be an important factor in determining the amplitude
of Babcock-Leighton-type flux transport dynamos.

\bibliographystyle{apj}
\bibliography{SFTC}

\clearpage

\begin{deluxetable}{ccccccc}
\tablewidth{0pc} \tablecaption{Dependence of
the polar field given in G on the  parameters of the flow perturbation}
\tablehead{\colhead{}   &
\multicolumn{6}{c}{$\Delta\lambda_\upsilon$}\\
\cline{3-7}
\colhead{} & & \colhead{
$10^{\circ}$ } &  & \colhead{ $15^{\circ}$ } & &
\colhead{$20^{\circ}$} \\
\cline{3-7}
\colhead{$\upsilon_0 \rm{[m\,s^{-1}]}$} & & &  & & &}
\startdata
0.0 & & 6.76 & & 6.76 & & 6.76 \\
2.5 & & 6.37 ($-6$\%) & & 6.08 ($-10$\%) & & 5.94 ($-12$\%) \\
5.0 & & 6.04 ($-10$\%) & & 5.52 ($-18$\%) & & 5.23 ($-23$\%) \\
10.0& & 5.56 ($-18$\%) & & 4.74 ($-30$\%) & & 4.15 ($-39$\%) \\
\enddata
\end{deluxetable}

\begin{deluxetable}{cccccccccc}
\tablecolumns{8} \tablewidth{0pc} \tablecaption{Dependence of
the polar field given in G on the activity belt parameters}
\tablehead{ \colhead{} & \colhead{} & \multicolumn{2}{c}{$\lambda_0=35^\circ$} & \colhead{}   &
\multicolumn{2}{c}{$\lambda_0=25^\circ$}& \colhead{}   &
\multicolumn{2}{c}{$\lambda_0=15^\circ$}\\
\cline{3-4} \cline{6-7} \cline{9-10}
\colhead{\ \ \ $\upsilon_0\rm{[m\,s^{-1}]}$} & & \colhead{0} &
\colhead{5} &   & \colhead{0} &
\colhead{5} & & \colhead{0}   &
\colhead{5} \\
\cline{1-1} \cline{3-4} \cline{6-7} \cline{9-10}
\colhead{{\hskip -12mm}$\Delta\lambda_{B}$} & &  & &   & &  & &  &}
\startdata
${\hskip -12mm}0^{\circ}$ & & 6.32 & 4.53 ($-28$\%) &  & 7.14 & 4.92 ($-31$\%) &  & 7.78 & 5.76 ($-25$\%) \\
${\hskip -12mm}6^{\circ}$ & & 6.25 & 5.26 ($-15$\%) &  & 6.76 & 5.52 ($-18$\%) &  & 6.86 & 5.75 ($-16$\%) \\
${\hskip -12mm}12^{\circ}$& & 6.23 & 5.87 ($-6 $\%) &  & 5.92 & 5.41 ($-8 $\%) &  & 6.28 & 5.98 ($-5 $\%) \\
\enddata
\end{deluxetable}

\clearpage

\begin{figure*}
  \centering
  \resizebox{\hsize}{!}{\includegraphics{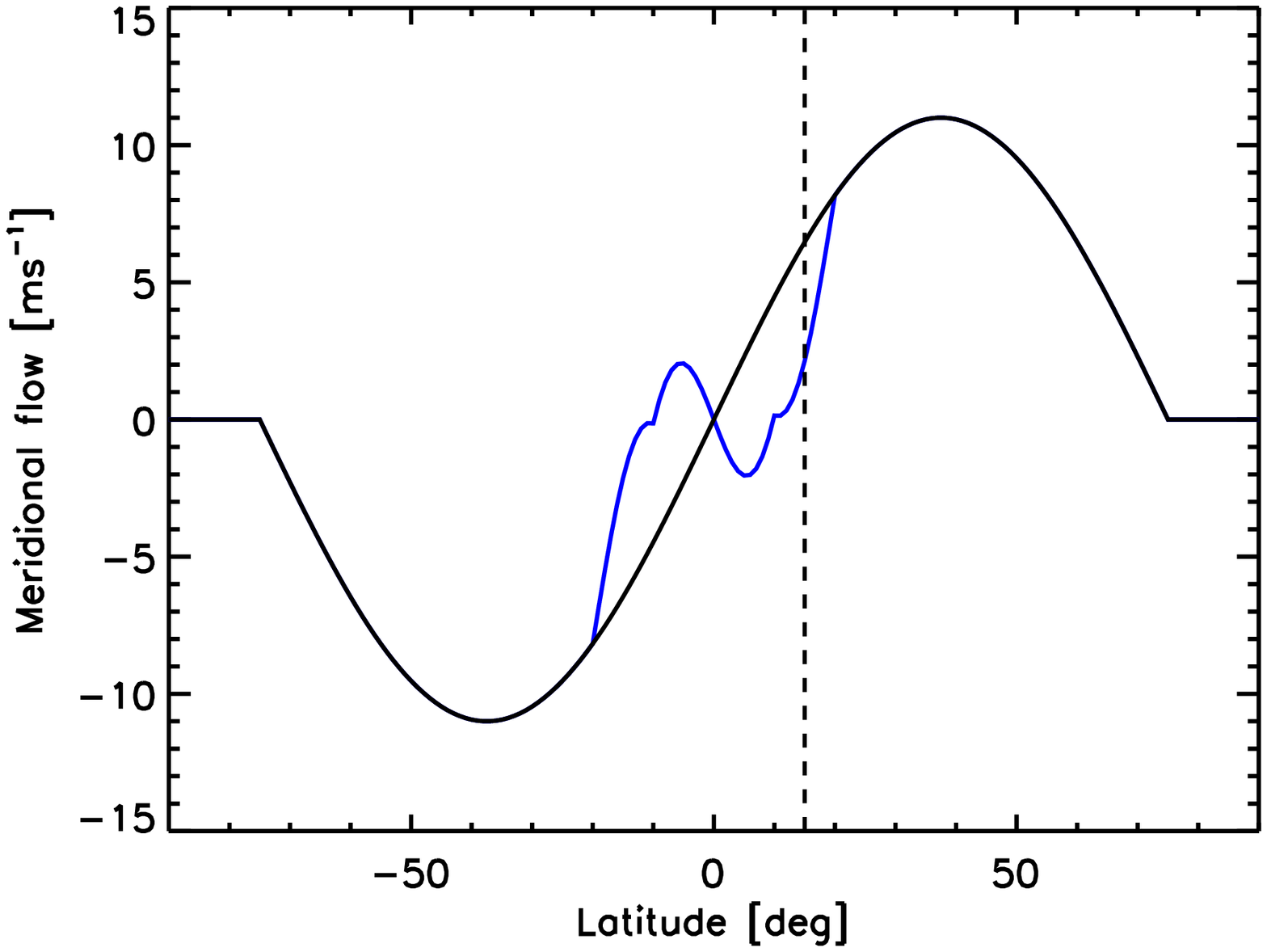}
                        \includegraphics{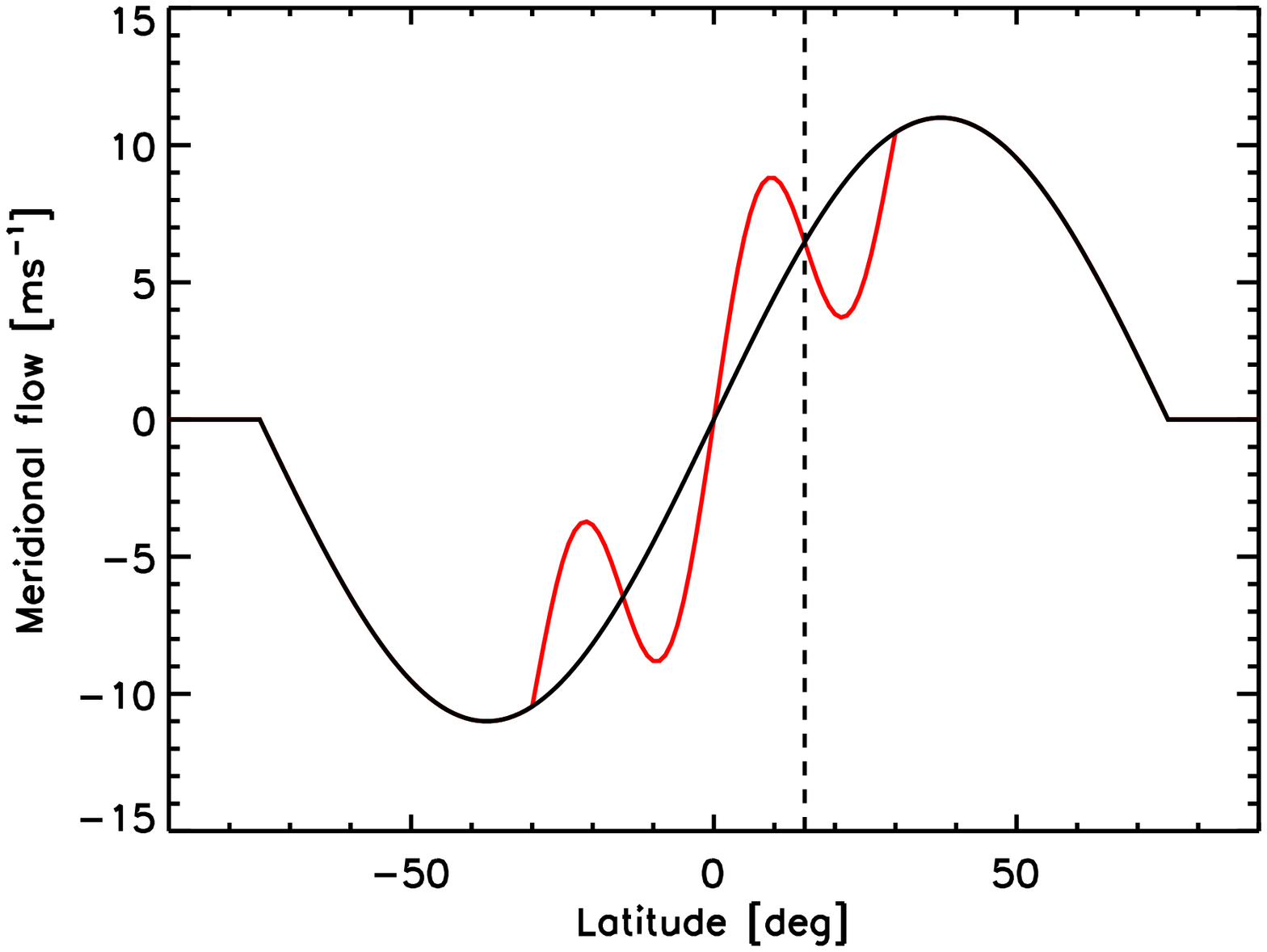}
                        \includegraphics{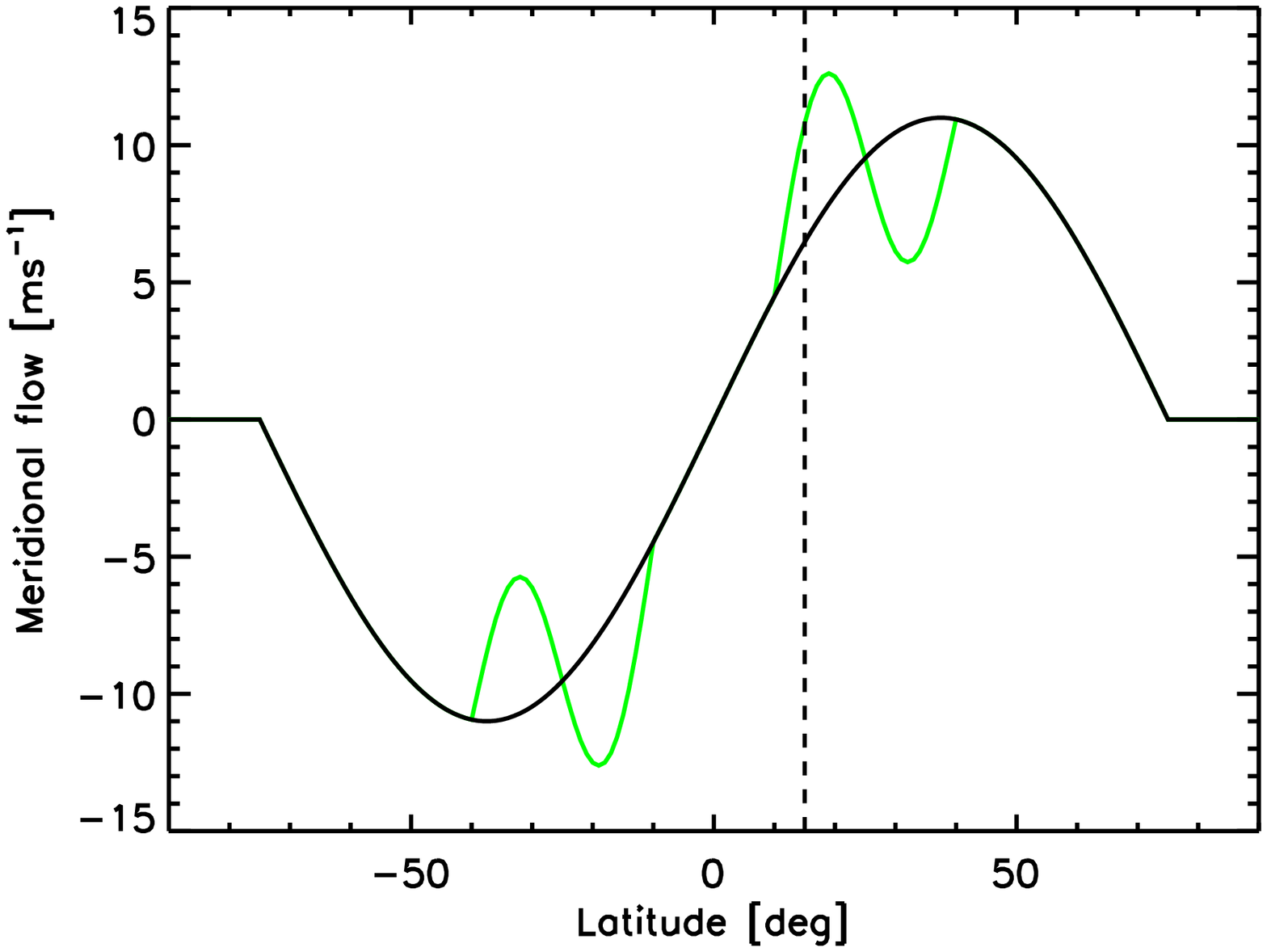}}
\caption{Meridional flow profiles considered for the simulations of a
  single bipole in a time-independent flow. Positive flow velocities are
  directed northward. The black curves in all panels show the profile
  without flow perturbation ($v_0=0$). The coloured curves represent the
  perturbed flow profiles for $v_0=5$~m$\,$s$^{-1}$,
  $\Delta\lambda_v=15^\circ$, and different central latitudes,
  $\lambda_c$, of the flow perturbation. {\em Left panel:}
  $\lambda_c=\pm 5^\circ$ (blue curve, coinciding with the black curve
  poleward of $\pm 20^\circ$); {\em Middle panel:} $\lambda_c=\pm
  15^\circ$ (red curve, coinciding with the black curve poleward of $\pm
  30^\circ$); {\em Right panel:} $\lambda_c=\pm 25^\circ$ (green curve,
  coinciding with the black curve poleward of $\pm 40^\circ$ and between
  $-10^\circ$ and $+10^\circ$). The dashed vertical lines indicate the
  mean latitude at which the bipolar region considered in
  Figure~\ref{fig:bip_snaps} is initiated.}
\label{fig:flow}
\end{figure*}

\begin{figure}
\epsscale{1.0}
\plotone{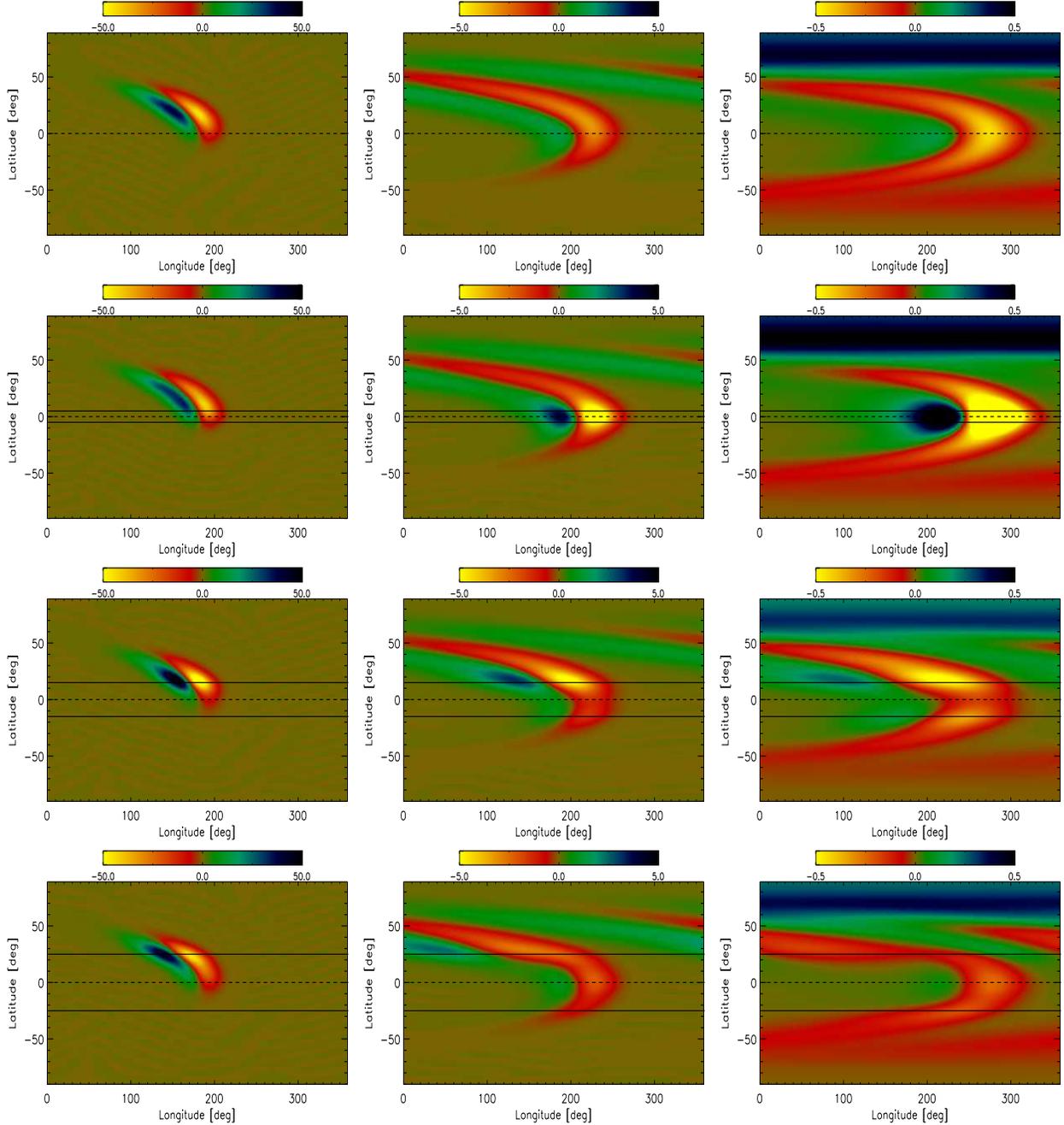}
\caption{Evolution of the magnetic field distribution resulting from a
single BMR with an initial unsigned flux of $1.9\,10^{23}$~Mx emerging
at $t=0$ at a mean latitude of $15^\circ$ and with a tilt angle of
$5.4^\circ$.  Each row shows three snapshots (after 4, 12, and 24
Carrington rotations, respectively). The top row corresponds to the case
without meridional flow perturbation ($v_0=0$, black curve in
Figure~\ref{fig:flow}). The rows below show the cases with perturbed
flow ($v_0=5$~m$\,$s$^{-1}$, $\Delta\lambda_v=15^\circ$) centered at
different latitudes: $\lambda_c=\pm5^\circ$ (second row, blue curve in
Figure~\ref{fig:flow}), $\lambda_c=\pm15^\circ$ (third row, red curve in
Figure~\ref{fig:flow}), and $\lambda_c=\pm25^\circ$ (bottom row, green
curve in Figure~\ref{fig:flow}). The dashed lines indicate the equator
while the full lines denote the central latitudes, $\pm\lambda_c$, of the
flow perturbation.  Magnetic field strengths are given in G. The
full time evolution for these cases can be viewed with aid of the
animations provided online as supplementary material.
}
\label{fig:bip_snaps}
\end{figure}

\begin{figure}
\epsscale{1.0}
\plotone{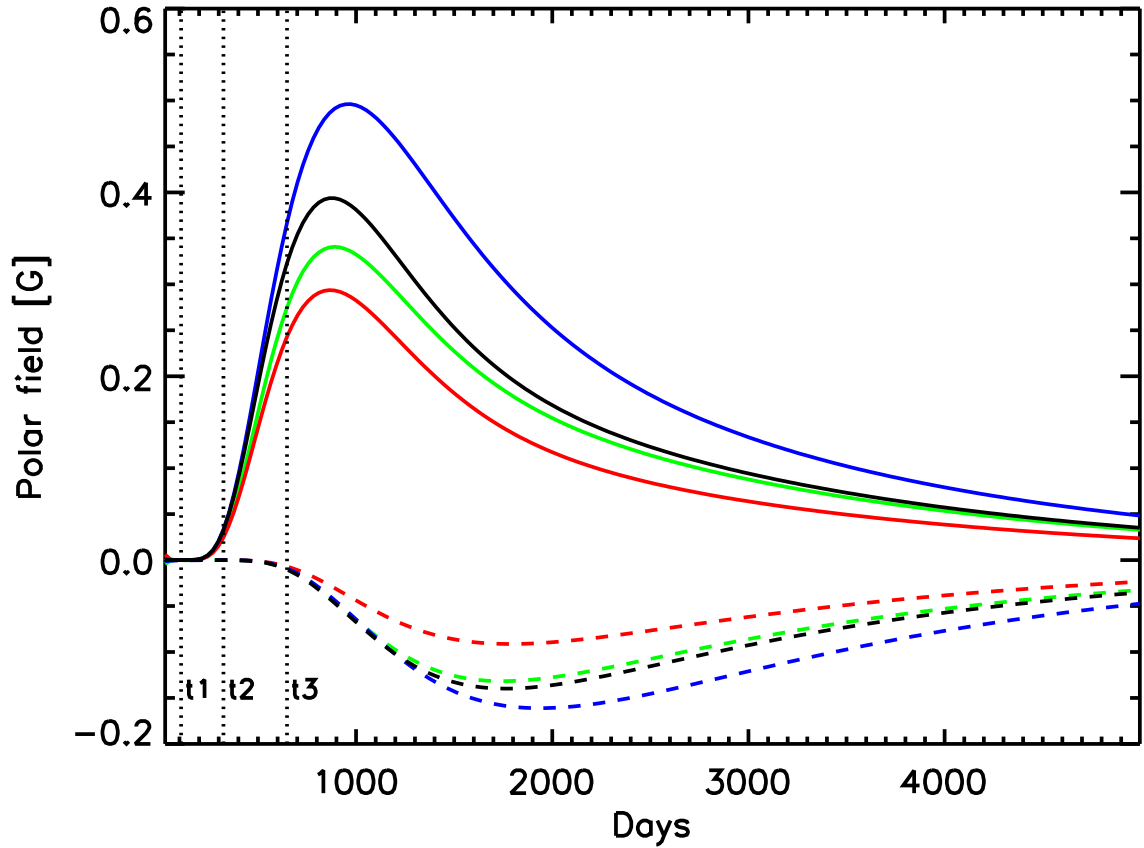}
\caption{Temporal development of the polar cap field (average radial
field poleward of $\pm 75^\circ$ latitude) for the simulations of single
BMRs (solid lines: North, dashed lines: South) shown in
Figure~\ref{fig:bip_snaps}.  The colors of the curves correspond to the
four flow patterns given in Figure~\ref{fig:flow}. The dotted vertical
lines indicate the times ($t_1,t_2,t_3$) corresponding to the snapshots
given in Figure~\ref{fig:bip_snaps}.}
\label{fig:bip_polar}
\end{figure}

\begin{figure}
\epsscale{1.0}
\plotone{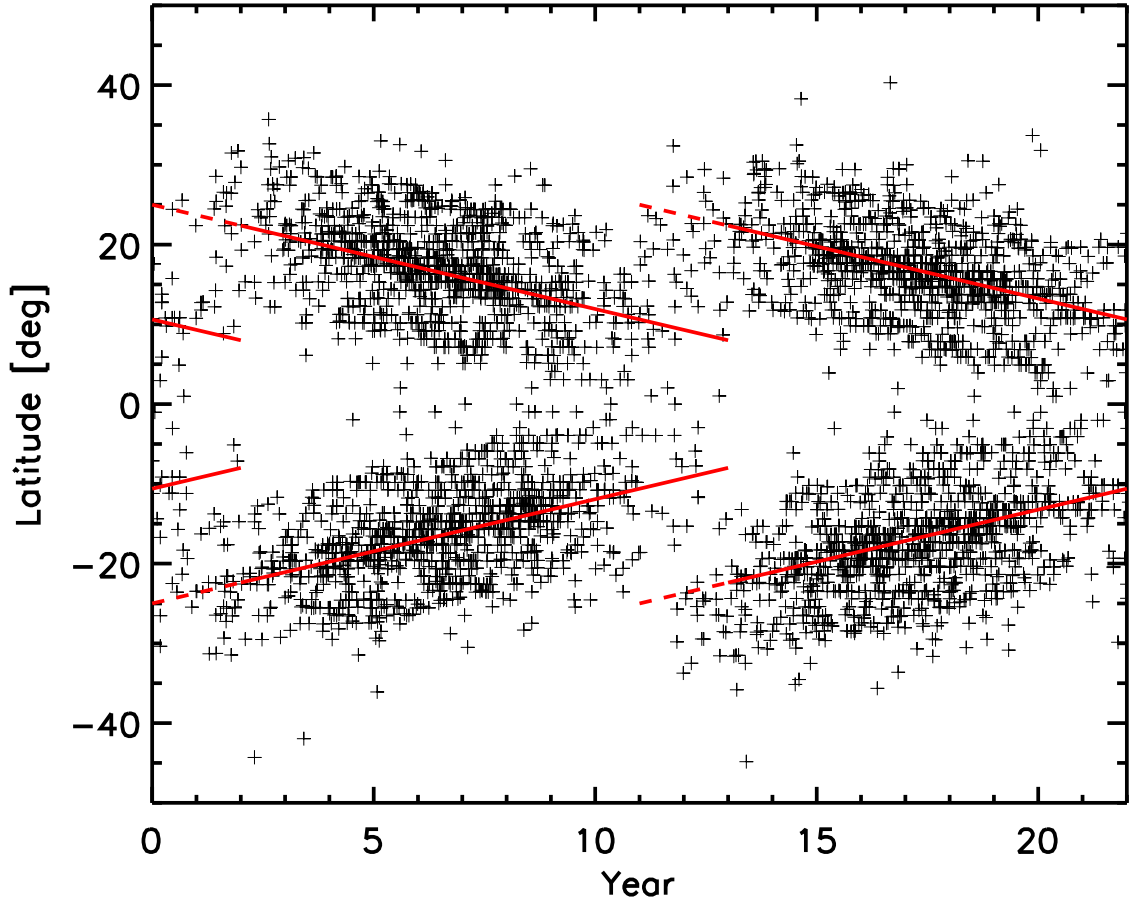}
\caption{Time-latitude diagram of new BMRs (+ symbols) used as input for the
flux transport simulation of solar cycles.  The red lines indicate the
central latitudes, $\lambda_\pm$, of the activity belts. The full parts of
the red lines also give the centers of the bands of meridional flow
perturbation (beginning 2 years after the start of the corresponding
activity belt).}
\label{fig:bf}
\end{figure}

\begin{figure}
\epsscale{1.0}
\plotone{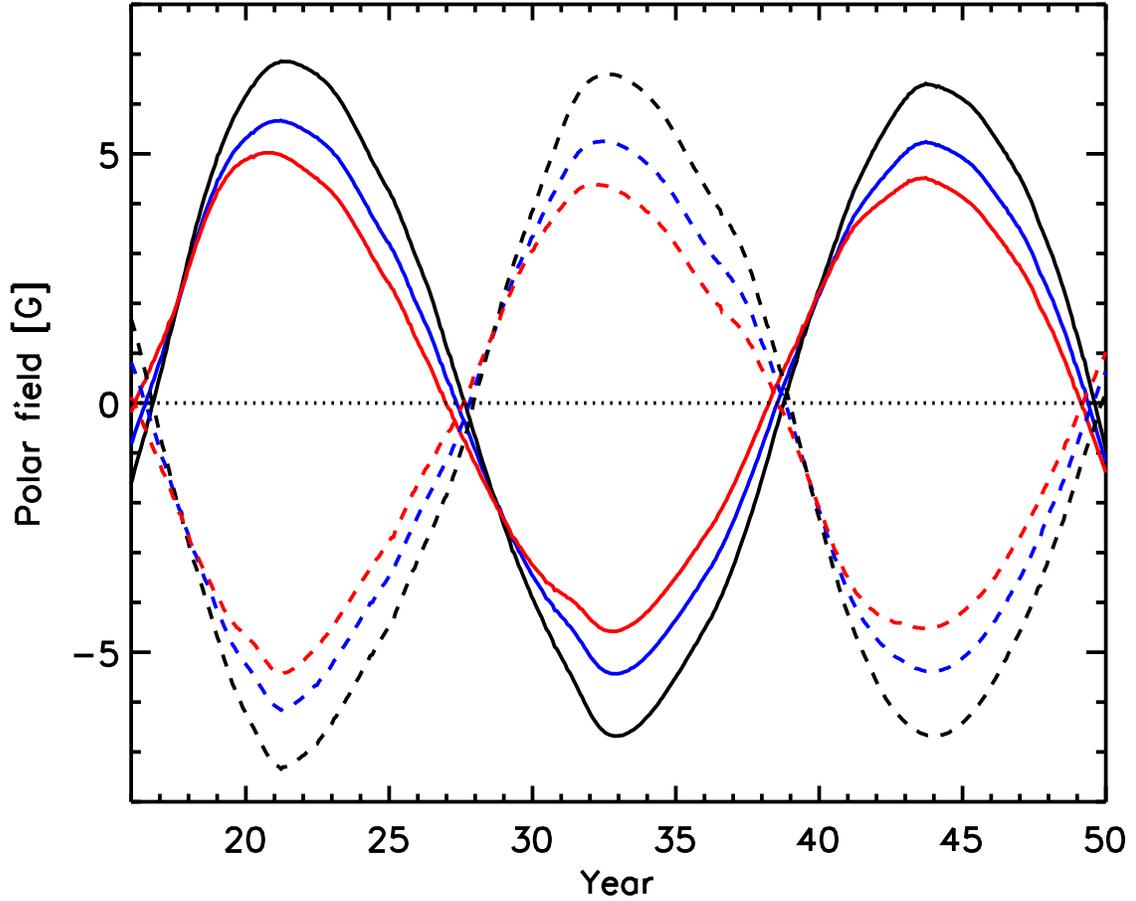}
\caption{Time evolution of the polar field strength (average radial
field poleward of $\pm 75^\circ$ latitude) for different values of the
perturbation amplitude of the meridional flow: $v_0=0$ (black curve),
$v_0=5$~m$\,$s$^{-1}$ (blue curve), $v_0=10$~m$\,$s$^{-1}$ (red curve).
The width of the bands of perturbed flow was taken as
$\Delta\lambda_v=15^\circ$.}
\label{fig:flow_amp}
\end{figure}

\end{document}